\documentclass[conference]{IEEEtran}

\IEEEoverridecommandlockouts
\usepackage{color}   
\usepackage{xcolor}
\usepackage[font=small]{subfig}
\usepackage{url}
\usepackage{hyperref}
\usepackage[english]{babel}
\usepackage[utf8]{inputenc}
\usepackage{blindtext}
\usepackage{listings}

\colorlet{punct}{red!60!black}
\definecolor{background}{HTML}{EEEEEE}
\definecolor{delim}{RGB}{20,105,176}
\colorlet{numb}{magenta!60!black}

\lstdefinelanguage{json}{
    basicstyle=\scriptsize\ttfamily, 
    numbers=left,
    numberstyle=\scriptsize,
    stepnumber=1,
    xleftmargin=2.5em,  
    numbersep=8pt,
    showstringspaces=false,
    breaklines=true,
    frame=lines,
    backgroundcolor=\color{background},
    literate=
     *{0}{{{\color{numb}0}}}{1}
      {1}{{{\color{numb}1}}}{1}
      {2}{{{\color{numb}2}}}{1}
      {3}{{{\color{numb}3}}}{1}
      {4}{{{\color{numb}4}}}{1}
      {5}{{{\color{numb}5}}}{1}
      {6}{{{\color{numb}6}}}{1}
      {7}{{{\color{numb}7}}}{1}
      {8}{{{\color{numb}8}}}{1}
      {9}{{{\color{numb}9}}}{1}
      {:}{{{\color{punct}{:}}}}{1}
      {,}{{{\color{punct}{,}}}}{1}
      {\{}{{{\color{delim}{\{}}}}{1}
      {\}}{{{\color{delim}{\}}}}}{1}
      {[}{{{\color{delim}{[}}}}{1}
      {]}{{{\color{delim}{]}}}}{1},
}

\newcommand{\ipnd}{\textsc{IPN-d}\xspace}
\newcommand{\ipnv}{\textsc{IPN-v}\xspace}

\definecolor{darkblue}{rgb}{0, 0, 0.5}

\usepackage{todonotes}

\begin{document}

\title{\ipnv: The Interplanetary Network Visualizer}

\author{
\IEEEauthorblockN{
Alice Le Bihan\IEEEauthorrefmark{1},
Juan A. Fraire\IEEEauthorrefmark{2}\IEEEauthorrefmark{3},
Pierre Francois\IEEEauthorrefmark{1},
Felix Flentge\IEEEauthorrefmark{4}
}
\IEEEauthorblockA{
\IEEEauthorrefmark{1}INSA Lyon, CITI, UR3720, 69621 Villeurbanne, France\\
\IEEEauthorrefmark{2}Inria, INSA Lyon, CITI, UR3720, 69621 Villeurbanne, France\\
\IEEEauthorrefmark{3}CONICET - Universidad Nacional de Córdoba, Córdoba, Argentina\\
\IEEEauthorrefmark{4}Ground Systems Engineering and Innovation Department, ESA/ESOC, Darmstadt, Germany 
}
}

\maketitle

\begin{abstract}
The Interplanetary Network (IPN) emerges as the backbone for communication between various spacecraft and satellites orbiting distant celestial bodies. 
This paper introduces the Interplanetary Network Visualizer (\ipnv), a software platform that integrates interplanetary communications planning support, education, and outreach. 
\ipnv bridges the gap between the complexities of astrodynamics and network engineering by enabling the generation and assessment of dynamic, realistic network topologies that encapsulate the inherent challenges of space communication, such as time-evolving latencies and planetary occlusions.
Leveraging the power of Unity 3D and C\#, \ipnv provides a user-friendly 3D interface for the interactive visualization of interplanetary networks, incorporating contact tracing models to represent line-of-sight communication constraints accurately. 
\ipnv supports importing and exporting contact plans compatible with established space communication standards, including NASA's ION and HDTN formats.
This paper delineates the conception, architecture, and operational framework of \ipnv while evaluating its performance metrics. 
\end{abstract}


\section{Introduction}
\label{sec_introduction}

Interplanetary Networks (IPNs) are crucial for enabling communication between spacecraft and satellites across planets, a necessity in the face of increasing interplanetary missions~\cite{alhilal2019sky}.
The Solar System Internet (SSI) term is commonly used to describe IPNs when applied explicitly within our solar system~\cite{hylton2022new}.
Central to the challenges of interplanetary networking are the significant communication \textit{delays} resulting from vast interplanetary distances, the signal \textit{disruptions} caused by celestial bodies occluding transmission paths, and the operational constraints of deep space equipment duty-cycling to conserve scarce energy resources.

Initially, the Delay-Tolerant Networking (DTN) architecture and the Bundle Protocol (BP) were developed as solutions to enable data transport over the challenging space environment of IPNs~\cite{RFC4838, RFC9171}. 
The core strategy of DTN lies in minimizing the assumption of immediate feedback from remote nodes and implementing temporal storage space to keep in-transit data until subsequent links in the path become available.

While substantial efforts have been invested in developing and assessing DTN and deep-space-related protocol stacks~\cite{ION-DTN, Burleigh2007, ione_sourceforge, HDTN, hylton2019delay, DTNME, cfs_goddard_2023, caini2023unibo, schildt2011ibr, uD3TN}, there is a pressing requirement for a tool that generates and visualizes authentic, extensive network topologies and sheds light on the dynamics of IPN topologies. 
Addressing the interdisciplinary gap, this tool must bridge the divide between the intricate celestial mechanics familiar to astronomers and the network design principles customary to software engineers, harmonizing both domains' expertise in advancing interplanetary communication.

Enhanced realism is crucial for network scenarios traditionally relying on manually configured, static contact topologies. 
This approach fails to account for the disruption dynamics and variations in data transmission rates and error probabilities as the communication angle changes, affecting the signal's passage through planetary atmospheres. 
While latency shifts might be minimal and less impactful during a single communication session, the fluctuating data rates and error rates during Earth-bound transmissions present a significant aspect of the complexity inherent in interplanetary networks. 

These critical modeling factors are already part of computational tools such as Ansys' STK~\cite{ansys_stk} or ESA's Godot~\cite{ESA2023Godot}, reflecting these intricate operational environments of space communication systems. 
However, they are conceived for accuracy and not intuitive apprehension of the effect.
Conversely, while proficient in certain aspects, visualization alternatives like Cesium~\cite{Cesium} and NASA Eyes~\cite{NASAEyes} do not adequately address the specific needs of the interplanetary domain or the intricate aspects of networking. 
Integrating these models with a visual tool would offer invaluable insights, enhancing an intuitive understanding of these complex interplanetary networks' dynamic, time-evolving nature.


\begin{figure*}
    \centering
    \includegraphics[width=1\linewidth]{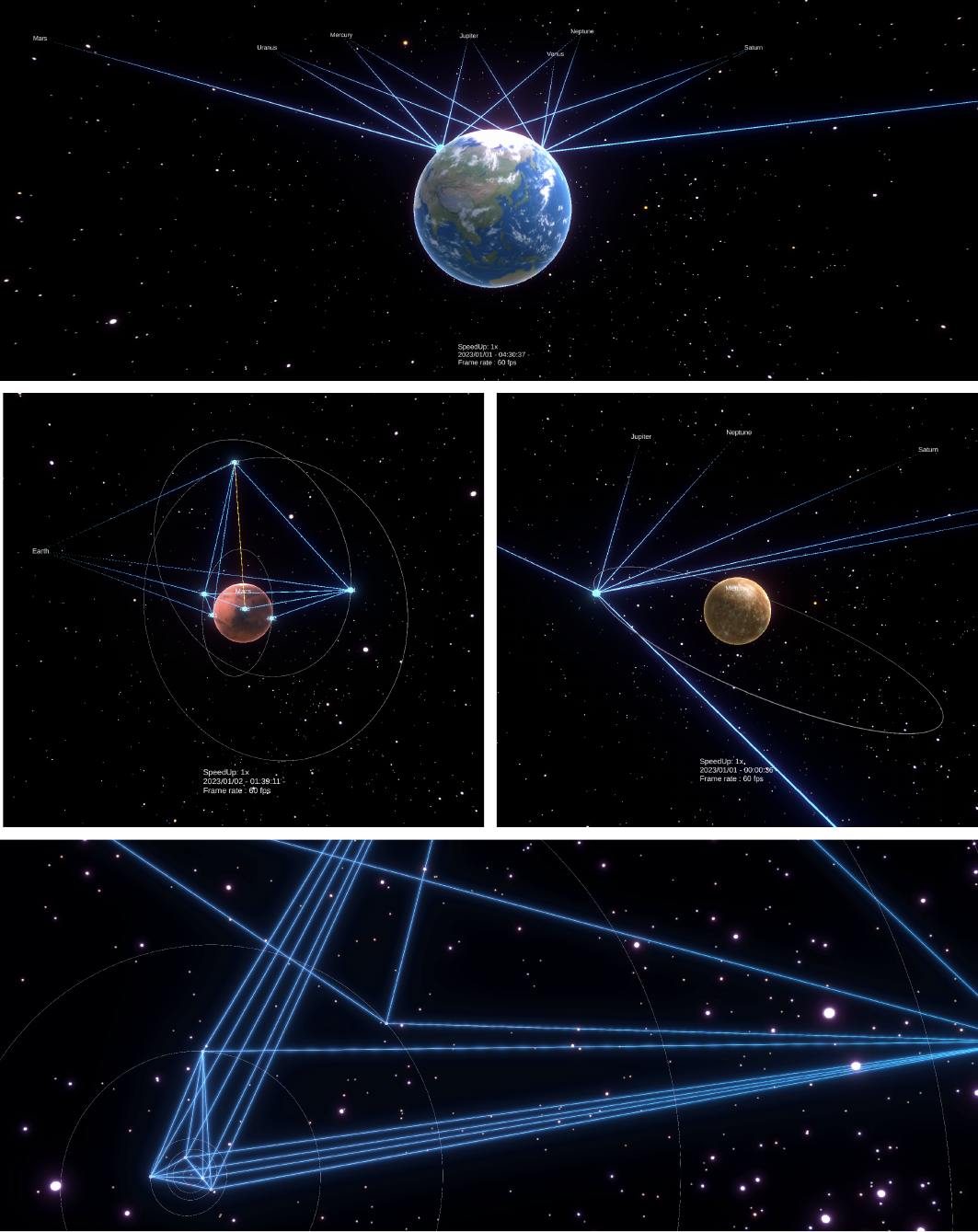}
    \caption{Screenshots of scenarios generated with \ipnv (1/2).}
    \label{fig:pics1}
\end{figure*}

\begin{figure*}
    \centering
    \includegraphics[width=1\linewidth]{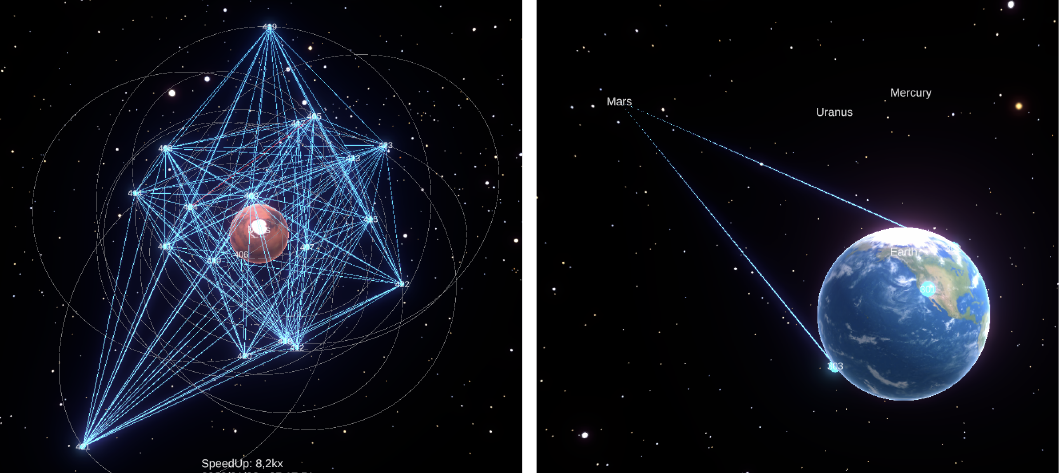}
    \caption{Screenshots of scenarios generated with \ipnv (2/2).}
    \label{fig:pics2}
\end{figure*}

This paper introduces the “Interplanetary Network Visualizer” (\ipnv).
\ipnv is a software developed to facilitate the \textit{planning}, \textit{education}, and \textit{outreach} efforts for projects involving communication within the solar system\footnote{The \ipnv tool is accessible on request. Usage guidelines, example files, and the Unity project, including the website code, are available in~\cite{ipnvRepo}.}.
The conception of the \ipnv described in this paper traces its roots to a 100-year projection video\footnote{The IPNSIG (\url{https://www.ipnsig.org/}) 100-year projection video can be accessed at \url{https://youtu.be/5rnbRdkrn70?si=eLiojnDE2jC7H62x}.} crafted by the authors for the InterPlanetary Networking Special Interest Group (IPNSIG). 
An early rendition of the \ipnv was utilized to produce this video, showcasing the tool's capabilities in visualizing interplanetary networks. 

Scenarios can be generated using \ipnv's built-in compute module based on simple models or by implementing export functions in existing space dynamics and planning software such as STK~\cite{ansys_stk} or ESA's Godot~\cite{ESA2023Godot} to create input files matching \ipnv's interface. 
Scenario elements within \ipnv's compute module include landers and orbiters as nodes based on Keplerian models implemented to mimic actual space dynamics. 
The compute module allows the export of realistic contact plans, following standards such as NASA's ION~\cite{Burleigh2007} and HDTN~\cite{hylton2019delay} formats.

Built using Unity 3D and programmed with C\#, \ipnv stands out for its user interface that enables users to interact with accurate models of interplanetary topologies and visualize dynamic, eye-catching 3D scenarios that represent the complex environment of space communication (see Figures~\ref{fig:pics1} and~\ref{fig:pics2}). 
Additionally, the \ipnv offers a time navigation element that can be adjusted live for speed, allowing scenarios to be accelerated or slowed down depending on the specific use case. 
Thus, it is a versatile educational and planning resource for the small yet focused research community engaged in interplanetary communication.

The remainder of this paper is organized as follows.
Section~\ref{sec_background} provides the motivation and context on which \ipnv was conceived.
Section~\ref{sec_ipnv} presents the \ipnv tool architecture, interfaces, models, and implementation details.
Section~\ref{sec_evaluation} discusses the main performance metrics of \ipnv.
Section~\ref{sec_conclusions} provides the outlook of this work and lists future development plans.

\section{Background}
\label{sec_background}

\subsection{Interplanetary Networking}

The historical milestones of space exploration, from NASA's launch of the Voyager missions to the deployment of several Mars rovers and orbiters, have underpinned the necessity for a robust and interoperable communication infrastructure~\cite{alhilal2019sky}.
In the context of the ``New Space", a novel interaction between the private and public sectors is fostering a rapid increase of interest in the interplanetary sector~\cite{eriksson2023outsourcing}.

Private entities like SpaceX and Blue Origin, along with legislative advancements in countries such as Luxembourg, have further underscored the potential of space industries, particularly in exploration and resource acquisition~\cite{lewicki2013planetary, steffen2022explore}.
Recent space initiatives like NASA's Lunar Communication and Navigation System (LCNRS)~\cite{TEMPOLCRNS}, ESA's Moonlight~\cite{giordano2021moonlight, ESAMoonlight}, and the joint LunaNet Interoperability Specification between NASA, ESA, and JAXA~\cite{israel2020lunanet, giordano2023lunanet, NASALunanet, NASA2023LunanetSpec} are a testament to the growing demand for IPN solutions. 
The Interagency Operations Advisory Group (IOAG) has identified interplanetary networking needs in its Lunar and Mars Communication Architectures~\cite{IOAG2023LunarComm}. 
It has selected the Bundle Protocol as the overall networking protocol, positioning DTN as the frontrunner architecture for the IPN.

\subsection{Delay Tolerant Networking}
\label{sec_DTN}

Delay Tolerant Networking~\cite{RFC4838} based on Bundle Protocol~\cite{RFC9171} rendered several notable implementations arguing in favor of the interest in IPN. 
Interplanetary Overlay Network (ION)~\cite{ION-DTN, Burleigh2007}, developed by NASA's JPL, is a comprehensive stack.
IONE~\cite{ione_sourceforge} mirrors ION but introduces experimental features. 
High-Speed DTN (HDTN)~\cite{HDTN, hylton2019delay}, from NASA's Glenn Research Center, focuses on high-performance networking. 
DTNME~\cite{DTNME} and 
{CFS}~\cite{cfs_goddard_2023}, from NASA Marshall and Goddard Space Flight Centers, respectively, contribute to flight grade stacks. 
{Unibo}~\cite{caini2023unibo} is geared towards academic research in DTN protocols. {IBR}~\cite{schildt2011ibr} prioritizes interoperability and terrestrial applications. 
{µD3TN}~\cite{uD3TN}, an evolution of µPCN, is unique for being developed by a private company.
Moreover, adapting the IP stack is a recent trend posing questions about the utility of DTN efforts, with lively discussions and standardization still in progress at IETF and CCSDS~\cite{Blanchet2023Revisiting}.

However, DTN does not only build the basis for interoperable communication in the lunar and Mars contexts. 
It is also considered for Earth observation missions with very high downlink data rates, where the store-and-forward mechanism provided by the protocol offers an elegant solution to overcome issues related to divergent data rates on the space link (up to 7 Gbps for the next generation of Earth Observation missions) and the terrestrial links between payload processing centers and the ground stations (several 100s of Mbps). 

The Bundle Protocol and related underlying re-transmission protocols for enhanced reliability, such as the Licklider Transmission Protocol (LTP)~\cite{bisacchi2022multicolor} or the High-Performance Reliability Protocol (HPRP) currently specified by CCSDS~\cite{CCSDS}, also offer distinct advantages in the case of optical direct-to-Earth links. 

Optical direct-to-Earth links are much more affected by atmospheric and weather effects (e.g., clouds) than traditional links based on RF~\cite{tai2018mars}. 
The automatic routing of bundles towards their destination allows for more opportunistic contacts in contrast to the currently pre-planned and explicitly initiated links for RF-based ground station contacts.

To unlock data handling across diverse optical and RF technologies, a networked communication approach where space assets can be reached through different routes provides additional resilience and robustness for monitoring and controlling heterogeneous space assets and enables the operation of such assets as a system compared to dedicated operations of individual assets.


In this context, a tool allowing the visualization of DTN scenarios is essential to assessing their behavior. 

\subsection{Related Toolings}

The landscape of space mission analysis tools is rich and varied, each offering unique features and limitations. 

\paragraph{STK (Systems Tool Kit)}
A widely recognized tool in the aerospace industry, STK offers robust OpenGL visualization capabilities, providing a high-quality graphical interface for simulating and analyzing space missions~\cite{stk}. Also, plugins such as the Contact Plan Designer were created to support DTN topology design~\cite{fraire2017introducing}. 
However, STK's primary limitation lies in its licensing model, which can hinder broader accessibility, especially for individual researchers or smaller organizations. 

\paragraph{Godot}
Developed by ESA/ESOC, Godot is geared towards orbit-related computations encompassing estimation, optimization, and analysis for various space missions~\cite{ESA2023Godot}. It stands out for its versatility and extensibility, catering to various mission types. Nevertheless, Godot is a library-based system requiring users to build their solutions using C++ or Python. Additionally, Godot's primary limitation is its lack of built-in visualization support, a significant drawback for tasks requiring graphical representations of mission scenarios.

\paragraph{Orekit}
This tool is known for its capabilities in orbit analysis, primarily focusing on accurate and efficient computations~\cite{Orekit}. Like Godot, Orekit is also library-based, providing minimal direct visualization support. While there is an associated application for visual representation, it is criticized for its lack of sophistication and limited features, which do not match the contemporary standards of graphical interfaces in mission analysis tools.

\paragraph{GMAT (General Mission Analysis Tool)}
As an open-source offering from NASA, GMAT is advantageous for its accessibility and collaborative development~\cite{GMATDocumentation}. It supports a range of mission design and analysis functionalities. However, GMAT's visualization capacities are relatively basic, particularly when illustrating complex link or network dynamics in space missions. 

Some online alternatives have shown remarkable user interfaces in the context of visualization tools.

\paragraph{NASA's Eyes}
This online visualization platform is a compelling, user-friendly tool that provides an immersive experience in space exploration~\cite{NASAEyes}. 
It allows users to explore and interact in real-time with a diverse range of celestial bodies, spacecraft, and the intricate dynamics of the solar system. 
The visualization goes beyond mere representation; it is an educational and engaging platform, offering in-depth insights into NASA's ongoing and future missions. 
However, the platform cannot be used for networking topology export or further analyses beyond the platform.

\paragraph{Cesium}
Cesium is a prominent geospatial platform known for its robust 3D mapping and visualization capabilities~\cite{Cesium}. 
Cesium renders high-resolution global imagery and terrain, making it an excellent fit for visualizing Earth orbit satellites. 
However, Cesium's primary focus on Earth-centric visualizations can be seen as a limitation, especially for endeavors requiring interplanetary or deep space perspectives. 
Additionally, Cesium's integration into broader workflows or toolsets is not seamless. Users often need to manually incorporate it into their existing systems, which could challenge those seeking an out-of-the-box solution. 

Existing astrodynamics modeling tools, while powerful, often have restrictive licensing models and lack sophisticated visualization capabilities. On the other hand, many current visualization tools, although visually impressive, fall short of providing detailed network export and analysis functionalities, especially in the interplanetary domain. 
In response to the identified gaps, our \ipnv emerges as a comprehensive solution designed to address these specific deficiencies in the context of IPN. 

\section{\ipnv}
\label{sec_ipnv}


Our \ipnv tool is deliberately crafted to bridge the identified gaps in existing toolings.
\ipnv offers a balanced integration of astrodynamics modeling interfaces with visualization capabilities tailored for Earth-centric and interplanetary missions. 
\ipnv is not constrained by licensing limitations, making it more accessible to a broader range of users, from individual researchers to large organizations. 
Moreover, \ipnv is equipped to visualize and analyze complex network topologies, a critical feature often overlooked in other platforms. 
This makes \ipnv a visualization tool, comprehensive mission analysis, and presentation suite.

The remainder of this section describes the architecture of \ipnv and some notable features. 

\subsection{Architecture}

\begin{figure}[]
    \centering
    \includegraphics[width=1.0\linewidth]{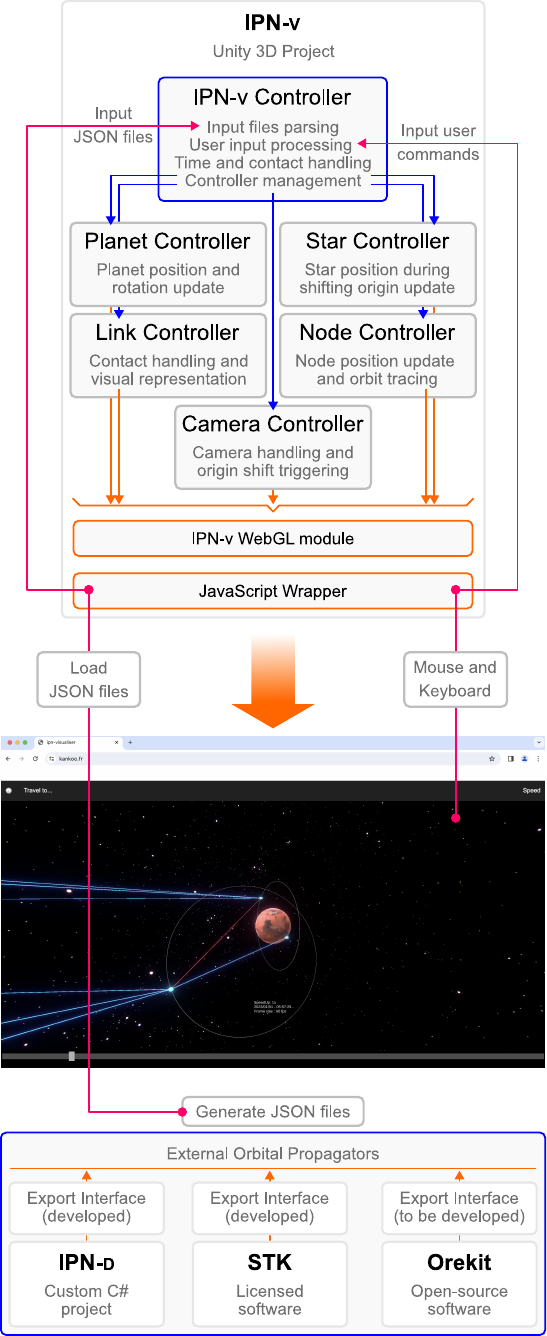}
    \caption{\ipnv Architecture.}
    \label{fig_architecture}
\end{figure}

Figure~\ref{fig_architecture} illustrates the architecture of \ipnv, developed within Unity 3D. 
The architecture is modular, with the \textit{\ipnv Controller} at its core, responsible for parsing input files, processing user inputs, and managing time and contact events within the network. 
It instantiates and delegates specific tasks to the \textit{Planet Controller}, which updates celestial bodies' positions and rotations, and the \textit{Star Controller}, which adjusts the central star position during an origin shift transformation explained below. 
The \textit{Link Controller} oversees the visual depiction of network connections, while the \textit{Node Controller} updates the positions of nodes and traces orbiters trajectories. 
The \textit{Camera Controller} manages camera control and viewpoint transitions. 
The visualization produced by these components is then rendered in the \ipnv WebGL module, which provides an interactive, web-based interface encapsulated by a JavaScript wrapper. 
This configuration allows for dynamic and comprehensive visualization of interplanetary network elements and their interactions over time, all within a user-friendly, browser-accessible environment.

\ipnv takes JSON configuration files as inputs and generates a 3D scene in which one can navigate, both in terms of position (from one planet to another) and time. 
The input files describe the general configuration of the simulation (celestial bodies in the scene, network nodes, start and end time), the contact plan, and the position of each object at every timestamp (with a step defined by the user). 
The \ipnv interface allows the visualization of topologies built using any DTN planning software through the input files. 

\subsubsection{\ipnd}
An IPN generation module, \ipnd, was implemented to exercise the tool for demonstration. 
It computes the scenario files from a defined start time to a simulation end time, progressing in defined steps. 
The module calculates all planet positions and rotations at each time frame using fundamental two-body Keplerian calculations, ensuring the dynamics of the celestial bodies are properly represented throughout the simulation. 
In parallel, it computes the positions of all nodes at each time increment, which is needed for assessing the network's connectivity.
The trajectory of orbiter nodes is also computed using two-body Keplerian mechanics.
Furthermore, \ipnd checks for line-of-sight (LOS) between all nodes at every step, identifying possible communication link opportunities or obstructions. 
Finally, \ipnd is responsible for storing all intermediary and final simulation data, which can be used for detailed analysis and visualization of the network's performance over time.
While all trajectories are calculated using Keplerian orbits, \ipnd offers a simple and efficient way to achieve low-accuracy position calculations, sufficient to evaluate the \ipnv visualization capabilities.

\subsubsection{Input files}
\label{sec_ipnv_files}

The \ipnd can generate the JSON input files for the visualization module. 
Alternatively, more sophisticated and accurate toolings, such as STK~\cite{ansys_stk} or ESA's Godot~\cite{ESA2023Godot}, can be extended to export \ipnv JSON files based on more accurate planets and orbiters propagators (e.g., SGP4 or HPOP).
Users should provide input files matching the \ipnv format to generate a scene correctly.  
Using those files, \ipnv generates the 3D rendering scene. 
A description of the structure of the files is provided below.

\paragraph{Configuration File}
A JSON {configuration file}, \texttt{config.json}, contains general information about the simulation, as shown in Listing~\ref{lst_config}. 

First, the JSON structure describes time. 
Users provide the start and end times of the scenario, which are defined in seconds since J2000 (January 1, 2000, 11:58:55.816 UTC). 
Users also specify the time step, which corresponds to the time in seconds between two instants where the positions of the objects are precisely described. 

Second, the file describes celestial bodies in the simulation. 
Users specify the stars and planets' names and radii (km).
Those do not necessarily need to be from the solar system, but if they aren't, they won't be textured, whereas a planet named "Mars" will automatically get a realistic texture.
This unlocks a high flexibility in \ipnv to visualize other systems beyond the Solar System.

Finally, \texttt{config.json} describes the network nodes. 
Planets have a list of nodes (orbiters or landers), each defined by a unique ID in the simulation and a name. Further work includes adding a list of Moons for each planet.

\begin{lstlisting}[language=json,caption=JSON configuration file structure, label=lst_config]
{
  "Time": {
    "SimulationStartTime": "seconds since J2000",
    "SimulationEndTime": "seconds since J2000",
    "Step": "step in seconds"
  },
  "Star": {
    "Name": "Star name",
    "Radius": "radius in km"
  },
  "Planets": [
    {
      "Name": "Planet name",
      "Radius": "radius in km",
      "Nodes": [
        {
          "ID": "node_1",
          "Name": "Node name",
        },
        {
          "ID": "node_2",
          "Name": "Node name",
        }
        // More nodes...
      ]
    }
    // More planets...
  ]
}
\end{lstlisting}

\paragraph{Contact Plan File}
A JSON {contact plan file}, \texttt{contactPlan.json}, lists every contact between nodes throughout the simulation, as shown in Listing~\ref{lst_contactplan}. 
Each contact is defined by the source node and end node IDs, and the start and end times (in seconds since J2000) of the contact. 
Additionally, a color can be specified to represent each contact. 
It is up to the user to decide what those colors mean: they could, for example,  highlight specific contacts, distinguish regions in space, link latency ranges, a path established by a routing strategy, or link properties as referred to in Section~\ref{sec_DTN}. In the future, colors will also be controllable from the user interface, for example to visualise on-going data transfers on specific links.

\begin{lstlisting}[language=json,caption=JSON contact plan file structure, label=lst_contactplan]
{
  "ContactPlan": [
    {
      "SourceID": "source_identifier",
      "DestinationID": "destination_identifier",
      "StartTime": "seconds since J2000",
      "EndTime": "seconds since J2000",
      "Color": [R,G,B]
    },
    {
      "SourceID": "source_identifier",
      "DestinationID": "destination_identifier",
      "StartTime": "seconds since J2000",
      "EndTime": "seconds since J2000",
      "Color": [R,G,B]
    }
    // More contacts...
  ]
}
\end{lstlisting}

\paragraph{Planet Files}
{Planet files} contain the list of each planet's known position and rotation throughout the simulation, as shown in Listing~\ref{lst_planetpositions}.  
One planet file is present for each planet listed in the configuration file. 
This list provides a discrete-time depiction of the planet movements, which \ipnv then uses to generate intermediate points using linear interpolations.
Each entry describes the planet's position (X, Y, Z) in kilometers and rotation (X, Y, Z) in degrees at a specific instant (in seconds since J2000). Planet coordinates are heliocentric.

\begin{lstlisting}[language=json,caption=JSON planet file structure, label=lst_planetpositions]
{
  "Positions": [
    {
      "Time": "Start Time",
      "PositionX": "km from star-center",
      "PositionY": "km from star-center",
      "PositionZ": "km from star-center",
      "RotationX": "degrees",
      "RotationY": "degrees",
      "RotationZ": "degrees"
    },
    {
      "Time": "Start Time + step",
      "PositionX": "km from star-center",
      "PositionY": "km from star-center",
      "PositionZ": "km from star-center",
      "RotationX": "degrees",
      "RotationY": "degrees",
      "RotationZ": "degrees"
    }
    // More position entries...
  ]
}
\end{lstlisting}

\paragraph{Node Files}
{Node files} describe man-made bodies such as satellites that are orbiting around or surface landers lying at the surface of a planet. 
As shown in Listing~\ref{lst_nodepositions}, such files have the same organization as \emph{planet files}, except that a node does not have a rotation parameter.
Furthermore, nodes' coordinates are expressed local to the planet to which the node is attached. 
\newline

\begin{lstlisting}[language=json,caption=JSON node file structure, label=lst_nodepositions]
{
  "Positions": [
    {
      "Time": "Start Time",
      "PositionX": "km from planet-center",
      "PositionY": "km from planet-center",
      "PositionZ": "km from planet-center"
    },
    {
      "Time": "Start Time + step",
      "PositionX": "km from planet-center",
      "PositionY": "km from planet-center",
      "PositionZ": "km from planet-center"
    }
    // More position entries...
  ]
}
\end{lstlisting}

\subsubsection{Web Hosting}
The visualization tool is hosted on a web server and uses WebGL technology. 
An interface developed in HTML/JavaScript/CSS allows users to travel through space and time and control the visualization speed.

\subsection{Notable features}

The tool relies on Double Precision, Origin Shift, and Contact Tracing to visually accurately observe a simulation state. We detail their use in the remainder of this section.  

\paragraph{Double Precision}
All calculations in \ipnd use double-precision floating points (float64) to account for the large distances between the bodies and avoid floating point errors. 
This type of error arises when the object is too far from the origin, and the number of bits available to represent its position is too low. 
With single-precision floating points (float32), only 6 to 9 digits are considered accurate, which is not high enough when working with 3D scenes at the scale of the solar system (Pluto is on average 5 900 000 000 km away from the sun). 
Positions become highly inaccurate, leading to unstable contact detection in the compute module. 
With double-precision floating points, 15 to 17 digits are considered accurate, which is enough for our needs. 

\begin{figure}
    \centering
    \includegraphics[width=1\linewidth, trim={18cm 1cm 1cm 2cm}, clip]{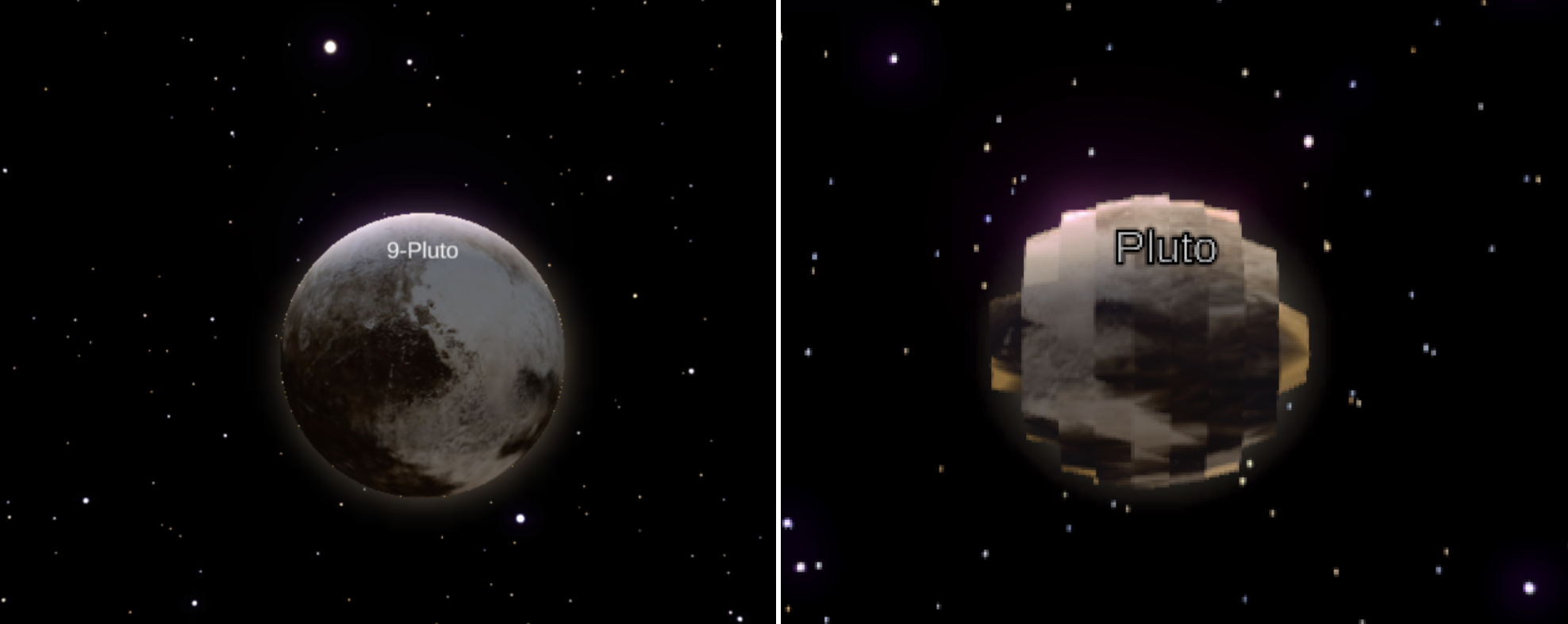}
    \vspace{-3mm} \\
    \includegraphics[width=1\linewidth, trim={2cm 1cm 18cm 3cm}, clip]{img/pluto_os.png}
    \caption{Screenshots of Pluto without (top) and with (bottom) origin shift in Unity.}
    \label{fig:origin_shift}
\end{figure}

\paragraph{Origin Shift}
For performance reasons, game engines like Unity rarely support float64; all positions are typically defined using Vector3 based on float32. 
Because double precision is not supported, rendering becomes inaccurate beyond a certain distance from the origin despite having precise positions calculated by \ipnd(see Figure \ref{fig:origin_shift}). 
A common strategy to build large worlds is to use a technique called origin shift. 
Instead of moving the camera toward a planet, keep the camera at the origin and move the planet toward it. 
In practice, anything visible by the camera will be close to the origin and, therefore, exempt from floating point errors. \ipnv leverages this technique paired with double-precision calculations to allow a smooth rendering of all objects and precise contacts, supporting a visually progressive transition from one origin to another. 

\paragraph{Contact Tracing}
The contacts between IPN nodes are governed by ray-casting models that account for the line-of-sight communication constraints.
To accurately depict a contact, the line representing a communication opportunity must propagate at the speed of light to capture the latency between transmission and reception. 
When a contact opportunity arises between A and B, anything sent by A will be received by B after a one-way light time. 
The line must emerge from A as long as it is detected that a transmitted bit has enough time to reach B before B gets out of sight. 
After this point, any transmission would be lost.
In this context, the \ipnv tool incorporates contact plan filtering to manage the complexity of network interactions.
Implementing this method of propagation is part of our ongoing work.


\section{Evaluation}
\label{sec_evaluation}

\begin{figure*}
    \centering
    \includegraphics[width=1\linewidth, trim={3cm 1.5cm 3cm 2cm}, clip]{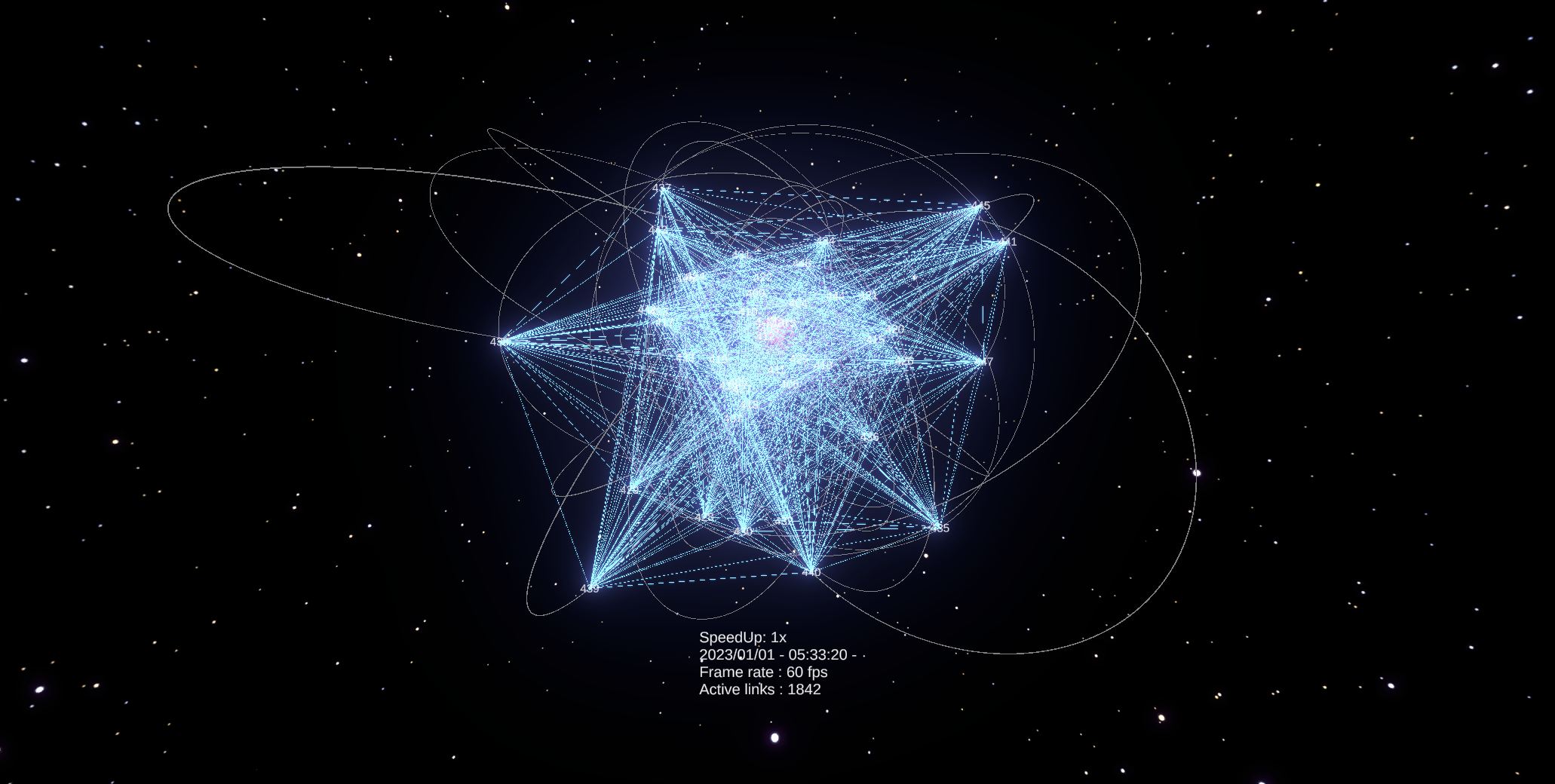}
    \caption{Screenshot of a scene with 1842 visible contacts around Mars.}
    \label{fig:1842Links}
\end{figure*}

We performed a performance analysis of the tool to assess its framerate and memory consumption under various simulation scenarios. 
The performance of \ipnv is evaluated with different numbers of contacts in the camera's field of view. 
All measurements were done in Chrome version 123.0.6312.58 on a 2020 M1 MacBook Air (RAM: 16GB, default screen resolution) using macOS Sonoma 14.2.1 (build 23C71). 
The Google Chrome Helper (GPU) process already uses 100~MB of RAM by simply opening Chrome without any website loaded. 

A 60~FPS framerate is observed up until 1900 visible contacts, with Google Chrome Helper (GPU) taking at most 425~MB of RAM. 
With 6100 contacts in the camera's FOV, it dropped to 25~FPS, using at most 520~MB of RAM. 
Finally, 16000 contacts led to a framerate of 12~FPS and a peak RAM usage of 790~MB, as reported by the GPU module of the Google Chrome Helper. 

Those experiments were performed under extreme cases where all contacts happened simultaneously around a single planet. 
It was unsurprising to notice that as soon as the links were outside the camera's FOV, the framerate peaked again at close to 60FPS, even with 16000 links being modeled.
Thus, it is not so much the total number of active contacts in the simulation that impacts performance as the total number of links rendered in front of the camera.

\section{Ongoing works}
\label{sec_ongoing}

This section outlines our planned works on our suite of tools, \ipnv and \ipnd. 
Our efforts aim to expand the visualization capabilities and enrich the scenarios that our tools can simulate. 
The near-term roadmap includes a set of strategic features:

\begin{enumerate}
    \item Development and validation of export functions compatible with STK and Godot, facilitating seamless integration through JSON file formats.
    \item Sophisticated modeling of nodal transitions based on orbital dynamics, such as orbit injection maneuvers, directly within node configuration files.
    \item Integration of GUI-based filtering to selectively visualize communication links, enhancing interactivity.
    \item GUI functionality for spotlighting specific DTN routes between nodes without manually adjusting the contact plan file.
    \item Simulation of contact establishment and fade-out to represent signal travel time, thus bringing a realistic temporal dimension to network animations.
    \item Establishment of a community-driven online database for sharing and disseminating scenarios.
    \item Transitioning the Unity component of our software to an open-source model with adequate licensing.
    \item Implementation of a contact plan designer within \ipnd, drawing inspiration from previous works, such as~\cite{fraire2017introducing}.
    \item Benchmarking \ipnd's performance and accuracy against industry standards such as STK and Godot.
    \item Development of an API that will allow users to interact with IPN-V.
\end{enumerate}

\section{Conclusions}
\label{sec_conclusions}

In concluding this presentation of the Interplanetary Network Visualizer (\ipnv), we reaffirm its significance as an innovative tool crafted to enhance our intuition of the dynamics and expanse of networking on an interplanetary scale. 
\ipnv emerges as a solution to address the intricate challenges of network visualization and Solar System Internet simulation. 
By integrating dynamic modeling and a visually intuitive interface, \ipnv achieved a synergy between astrodynamics and software engineering.

This work has successfully demonstrated \ipnv's potential as an open, extensible platform capable of adapting to diverse planetary system definitions and accessible through user-friendly web and file-based interfaces. 
Our commitment to the open-source community is expected to catalyze further developments, encouraging collaborations that will extend the frontiers of space networking. 

As we anticipate the open-source release of \ipnv, we envision it to be a cornerstone in the planning, education, and strategic dissemination of interplanetary networking knowledge, propelling the future of space exploration. 
In the long run, \ipnv aims at fostering the collective interest of researchers, educators, and enthusiasts to chart new research and development courses in interplanetary networking.

\section*{Acknowledgement}
This research has received support from the European Union's Horizon 2020 R\&D program under the Marie Skłodowska-Curie grant agreement No 101008233 (MISSION project) and the French National Research Agency (ANR) under the STEREO project ANR-22-CE25-0014-01.

\bibliography{references}
\bibliographystyle{IEEEtran}
\end{document}